\def \bc {\begin{center}}
\def \ec {\end{center}}

\def \bfr {\begin{flushright}}
\def \efr {\end{flushright}}

\def \v {\vskip}

\def \ba {\begin{array}}
\def \ea {\end{array}}

\def \bea {\begin{eqnarray}}
\def \eea {\end{eqnarray}}

\def \be {\begin{equation}}
\def \ee {\end{equation}}
\def \p {\partial}

\def \d {\hbox{d}\,}
\def \square {\hbox{$\sqcup\!\!\!\!\sqcap$}} 

\def \e {\hbox{e}}

\def \g {\bar{g}} 
\def \npdos {(\nabla\phi)^2}
\documentstyle[12pt,a4]{article}
\begin{document}
%
\thispagestyle{empty}
\hfil gr-qc/9702040\break
 
\vskip 10mm 
\begin{center} 

{\bf GENERALIZED SYMMETRIES AND 
INVARIANT MATTER COUPLINGS IN  
TWO-DIMENSIONAL DILATON GRAVITY}  
\footnote{Work partially supported by the 
{\it Comisi\'on Interministerial de Ciencia y Tecnolog\'{\i}a}\/ 
and {\it DGICYT}.}

\vskip5mm 
 Miguel Navarro$^{1,2}$\footnote{http://www.ugr.es/
$\widetilde{}$mnavarro;
 mnavarro@ugr.es} 

\end{center}

\bigskip%

\normalsize
\v2mm

{\it Instituto de Matem\'aticas y F\'\i sica Fundamental, 
        CSIC. Serrano 113-123, 28006 Madrid, Spain.}             

\centerline{and}

{\it Instituto Carlos I de F\'\i sica Te\'orica y Computacional,
        Facultad  de  Ciencias, Universidad de Granada. 
        Campus de Fuentenueva, 18002, Granada, Spain. }

\vskip10mm

\begin{center}
                        {\bf Abstract}
\end{center}

\footnotesize 

New features of the generalized symmetries of generic 
two-dimensional dilaton models of gravity are presented and  
invariant gravity-matter couplings are introduced. 
We show that there is a continuum set of Noether symmetries,  
which contains half a de Witt algebra.  Two of these symmetries are 
area-preserving transformations. We show that 
gravity-matter couplings which are invariant under area preserving 
transformations only contribute to the dynamics 
of the dilaton-gravity sector with a reshaping of 
the dilaton potential. The interaction with matter by means of 
invariant metrics is also considered. We show in a constructive 
way that there are metrics which are invariant under two 
of the symmetries. The most general metrics and  
minimal couplings that fulfil this condition are found. 

\normalsize 
\vskip5mm

\newpage
\setcounter{page}{1}
\section{Introduction}

Currently one of the main objectives of 
Theoretical Physics is to devise a  quantum theory of gravity. 
The four-dimensional Einstein-Hilbert gravity
theory -- and, in general, four- and higher-dimensional models of 
gravity -- is unfortunately very complex to handle. 
Toy models which share their most 
relevant features with Einstein's gravity should, therefore, 
play an important role here. 2D dilaton models of gravity  
are general covariant models which  
in addition to the two-dimensional metric  
also involve a scalar (dilaton) field (for a review see 
Ref. \cite{[Giddings]}). When coupled to matter, these models have  
solutions describing the formation of
two-dimensional black holes and Hawking radiation 
\cite{[CGHS],[cadoni]}.  Moreover, and unlike 
their four-dimensional counterparts, these  models 
are renormalizable. Thus they may be an useful tool 
to explore the final fate of black holes and solve the 
information puzzle. 

Unfortunately, in spite of their being much 
simpler than their  higher-dimensional cousins, 
only few of them, notably the 
CGHS model, have been shown to 
be solvable when interacting with matter. This represents 
a serious drawback, as it is difficult to distinguish 
in the dynamics of these models which 
is due to a particular feature of the CGHS 
model and which corresponds 
to general properties of gravity.
These theories would, then, be far more useful 
if the developments which have been made with the CGHS model 
could be extended to more general ones,  especially 
spherically symmetric gravity. 
If this were the case, we could be more confident   
that the experience gained from 
these two-dimensional toy models would actually 
be useful in the four-dimensional case. 
Moreover, the quantum nature of the CGHS model remains elusive 
(see, for instance, Ref. \cite{[Jackiw2]}). Other solvable models   
might not face the difficulties which have been found when 
trying to quantize this model. 

As is well known, solvability is usually related to invariances
 -- this being the reason that classical solvability usually 
implies quantum solvability. New generalized symmetries 
(we use the adjective ``generalized'' 
because they involve derivatives of the fields) 
have been recently uncovered for generic dilaton 
gravity which generalize those of 
the CGHS model \cite{[symbh],[goslar]}.   
Therefore, it is natural to study whether or not these symmetries 
can be used to find invariant gravity-matter interactions, thereby  
providing solvable models. 

In the present paper, we shall consider two different 
approaches to introduce symmetric gravity-matter couplings. 
In the first approach  (Section 4),   
we consider couplings which are invariant under area-preserving 
transformations of the metric. In the second one (Sect. 5),  
we consider couplings which are 
constructed by means of an invariant metric.
For the first approach, we show 
that, with regard to the gravity sector, interaction   
with general area-preserving couplings simply amounts to a 
reshaping of the dilaton potential. For the second approach, we 
show that metrics and conformal couplings can be constructed 
which are invariant under two of the symmetries. The most general 
invariant metrics and conformal couplings are constructed. 
In the last section, we briefly discuss several natural continuations 
of the present developments.  

Firstly, the CGHS model and the generalized 
symmetries of the generic 2D dilaton models are briefly reviewed
and some new features are presented. 

\section{Two-dimensional dilaton gravities and the\break CGHS model} 

The generic models of two-dimensional dilaton gravity   
are defined by means of the action  

\be S_{{GDG}}\left(\widetilde{g},\widetilde{\phi}\right) 
= {1\over2\pi} \int d^2x\sqrt{-\widetilde{g}}
\left[ D(\widetilde{\phi}) \widetilde{R} + 
H(\widetilde{\phi})\left(\nabla\widetilde{\phi}\right)^2 + 
F(\widetilde{\phi})\right] 
- S_{M}\label{GDG0}\ee 
where $D,H$ and $F$ are arbitrary functions, and $S_M$ is the 
gravity-matter interaction term. 

A result which is particularly useful 
is that after suitable redefinitions of the two-dimensional metric 
$\widetilde{g}_{\mu\nu}\rightarrow{g}_{\mu\nu}$ 
and the dilaton field $\widetilde\phi\rightarrow\phi$,   
any action can be brought to the form \cite{[Banks],[Gegenberg]} 

\be \tilde{S}_{{GDG}}= S_V - S_M\label{GDG}\ee 
where 
\be S_V={1\over2\pi} \int d^2x\sqrt{-{g}}
\left({R}{\phi} + {V} ({\phi})\right) \label{SV} \ee 

The CGHS or 
string-inspired model of two-dimensional 
dilaton gravity \cite{[Witten],[CGHS]},  with action

\be S_{{CGHS}} = {1\over2\pi} \int d^2 x \sqrt{-g}
        \left[(R\phi  + 4\lambda^2 ) 
-{1\over2} (\nabla f)^2 \right]\label{CGHS}\ee 
has attracted particular attention because 
it is exactly solvable at the classical as well as the semiclassical 
levels \cite{[RST],[BPP]}.  
Solvability of this model is due to the existence of 
a sufficient number of free-field and Liouville 
equations, and this, in turn, is related to the existence of symmetries. 
Let us  for the moment restrict our attention to the model  
without cosmological constant, in which case the symmetries 
are particularly simple.  It is easy to see that, in addition to 
the familiar transformation of the matter sector 
 \be \delta_f \phi = 0,\quad 
\delta_f g_{\mu\nu}=0, \quad  \delta_f f=\epsilon\ee 
the following two transformations are also symmetries 
of the string-inspired model (\ref{CGHS}) with $\lambda=0$ 

\bea \delta_R \phi&=&\epsilon, \quad 
\delta_R g_{\mu\nu}=0, \quad  \delta_R f=0\\
\delta_\phi \phi&=&0,\quad \delta_\phi g_{\mu\nu}=
\epsilon g_{\mu\nu}, \quad  \delta_\phi f=0\eea
These symmetries correspond to the following free-field  
equations:

\be  \square f =0, \quad R=0, \quad\square \phi =0\label{eqcghs}\ee 
In turn, these free-fields equations imply classical solvability 
for the model. This can be easily seen by choosing the conformal 
gauge for the metric 

\be \d s^2=2\e^\rho\d x^+ \d x^-\label{conformalmetric}\ee
in terms of which we have $R= -2\e^{-\rho}\p_+\p_-\rho$ and  
$\square = 2\e^{-\rho}\p_+\p_-$. 

\section{New symmetries in generic 2D dilaton gravity}

Now, let us go back to the general Lagrangian in Eq. (\ref{GDG}).  
It can be shown that generic dilaton gravity 
without matter is highly symmetric, also \cite{[symbh],[goslar]}. 
This provides a symmetry-based explanation 
for the solvability of these models.  

For the Lagrangian in Eq. (\ref{GDG}) we have 

\bea 
\delta {\cal L} &=& \frac{\sqrt{-g}}{2\pi}
\left\{\vphantom{{1\over2}}
\left[ R + V'(\phi) - T_\phi\right]\delta \phi\right. \nonumber \\
&&+\left[g_{\mu\nu}\square\phi -\frac12g_{\mu\nu}V(\phi)
- \nabla_\mu\nabla_\nu\phi-T_{\mu\nu}\right]\delta g^{\mu\nu}\label{deltalag}\\
&&+\left(E-L\right)_A\delta f^A\nonumber\\
&&+ \nabla_\alpha\left[-\phi(g^{\mu\nu}\nabla^\alpha
\delta g_{\mu\nu}   - g^{\alpha\mu}\nabla^\nu\delta g_{\mu\nu})
   - \nabla^\alpha\phi \, g_{\mu\nu}\delta g^{\mu\nu} +
   \nabla_\nu\phi\, \delta g^{\nu\alpha}\right.\nonumber\\
&&+ \left.\left. j^\alpha_{M}\>\>\right]\right\}\nonumber\eea
Here $f^A$ are the matter fields, $\left(E-L\right)_A=0$ are the 
Euler-Lagrange equations of motions for these fields, and
 $j^\mu_{M}$ is the matter contribution to the symplectic 
potential current of the model. 

The equations of motions are therefore:
 
\bea 
R+ V'(\phi)&=&T_\phi\nonumber\\
g_{\mu\nu}\square\phi -\frac12g_{\mu\nu}V(\phi)
- \nabla_\mu\nabla_\nu\phi&=&T_{\mu\nu} \label{eqofmot}\\
\left({E-L}\right)_A&=&0\nonumber\> 
\eea 
In absence of matter, it is easy to show that the 
general action for the 2D dilaton models   
(\ref{GDG}) is invariant under the following symmetries (note 
the change in notation with 
respect to Refs. \cite{[symbh],[goslar]}):  

\bea 
\delta_a \phi=0 
&, & 
\delta_a g_{\mu\nu}= 
g_{\mu\nu}a_\sigma\nabla^\sigma\phi-\frac12 
\left( a_\mu \nabla_\nu\phi + a_\nu \nabla_\mu \phi \right)
\nonumber \\
\delta_1\phi=0 
&,& 
\delta_1 g_{\mu\nu}=
\epsilon_1\left({g_{\mu\nu}\over\left(\nabla\phi\right)^2}
-2{\nabla_{\mu}\phi\nabla_{\nu}\phi\over\left(\nabla\phi\right)^4}\right)
\\
\delta_2\phi = \epsilon_2
&,&
\delta_2 g_{\mu\nu}=\epsilon_2 
V\left({g_{\mu\nu}\over\left(\nabla\phi
\right)^2}-2{\nabla_{\mu}\phi\nabla_{\nu}\phi
        \over\left(\nabla\phi\right)^4}\right) 
\nonumber
\eea 
where $a_\mu$ is any arbitrary constant bivector. 

The Noether currents are, respectively,  

\be
J^{\mu\nu}=  g^{\mu\nu}E\>,\quad 
j_{1}^{\mu}={\nabla^{\mu}\phi\over\left(\nabla\phi\right)^2}
\>,\quad
 j_2^{\mu}=j_R^{\mu}+V{\nabla^{\mu}
\phi\over\left(\nabla\phi\right)^2}\label{currents}
\ee
where $E= \frac12\left((\nabla\phi)^2 -J(\phi)\right)$ with  
$J(\phi)$ a primitive of $V(\phi)$:  
$J^{\prime}\left(\phi\right)=V(\phi)$.  

 Now,  
$j^\mu_1$ and $j^\mu_2-j^\mu_R$ satisfy the integrability 
condition:

\be \epsilon^{\mu\nu}\nabla_{\mu}(j_1)_\nu=0
=\epsilon^{\mu\nu}\nabla_{\mu}(j_2-j_R)_\nu\ee
Thus, the conservation 
law for the currents $j_1^\mu$, $j_2^\mu$ turns out to imply the 
existence of two free fields. The free-field equations are, 
respectively: 

\bea 
\square j_1&=&0 \>, \nonumber\\ 
R +\square j_2 &=&0 \>,\label{ffields}
\eea 
where 
\be
j_1=\int^\phi {d\tau\over 2E+J\left(\tau\right)} 
\ee
and 
\be j_2 =\log(2E+J)
\ee 

Now, Eq. (\ref{ffields}) lead directly to the general solution to 
the equations of motion (\ref{eqofmot}). To show this,  
let us choose light-cone coordinates and fix the residual 
conformal gauge as follows 

\be j_1=\frac12(x^+ -x^-)\ee 
Then, the classical general solution of the theory 
can be (implicity) given as follows: 

\bea\int^\phi {d\tau\over 2E+J\left(\tau\right)}
&=&\frac12(x^+ -x^-)\nonumber\\
\rho&=&(2E+J(\phi))\eea 

Moreover, due to the peculiar conservation law for $E$ 

\be \p_\mu E = 0, \quad \mu=1,2\label{lce}\ee
if $j^\mu$ is a conserved current, so is $f(E)j^\mu$, for 
arbitrary function $f$.  In the particular case of $j_1^\mu$, 
these conserved currents turn out to be the Noether current
of local (generalized) symmetries 

\bea \delta_f\phi&=&0\>,\nonumber\\
\delta_f g_{\mu\nu}&=& -\epsilon f'(E)
\left(g_{\mu\nu}
-{\nabla_{\mu}\phi\nabla_{\nu}\phi\over\left(\nabla\phi\right)^2}
\right) + 
\epsilon f(E)\left({g_{\mu\nu}\over\left(\nabla\phi\right)^2}
-2{\nabla_{\mu}\phi\nabla_{\nu}\phi\over\left(\nabla\phi\right)^4}
\right)\label{deltafe}\nonumber\eea 
In particular, $\delta_E$ is given by:

\be 
\delta_E\phi=0 
\quad,\quad 
\delta_E g_{\mu\nu}= -\frac{\epsilon_3}2\left[g_{\mu\nu}+ 
J\left({g_{\mu\nu}\over\left(\nabla\phi\right)^2}
-2{\nabla_{\mu}\phi
\nabla_{\nu}\phi\over\left(\nabla\phi\right)^4}\right)\right]
\>. \label{deltaE}
\ee

These symmetries close the algebra 

\be [\delta_f,\>\delta_g]=\frac12\delta_{(f'g-g'f)} \label{alsym}\ee
If restricted to analytic functions, they define half a de Witt 
algebra 

\be [\delta_n, \delta_m] = \frac12\delta_{(n-m)} \label{deWitt}\ee 
In particular $\delta_1, \delta_E$ and $\delta_{E^2}$ 
close a $sl(2,R)$ algebra. 

In the limiting case $V=0$ the symmetry $\delta_2$ 
is the symmetry 
$\delta_R$ of the string-inspired model with no cosmological 
constant. Moreover, -2$\delta_E$  coincides with $\delta_\phi$. 
 It is apparent, therefore, 
that we have generalized the symmetries of the CGHS 
model to an arbitrary 2D dilaton gravity model. 

We should emphasize that the transformations 
just described are symmetries for 
generic dilaton models.  For particular potentials, these symmetries 
present special features. Notably, it turns out that, 
although for a generic potential none of these symmetries 
is conformal, 
for $V=4\lambda^2$ (the string-inspired model) or 
$V=4\lambda^2\e^{\beta\phi}$ (the so-called exponential model 
\cite{[symbh]}), a linear combination of these symmetries is conformal.  
These conformal symmetries are
$\delta_2 -4\lambda^2\delta_1$ 
and $\delta_2 +2\beta \delta_E$, respectively. 
The coupling to conformal matter therefore 
preserves this symmetry.  
However, these two models 
are the only ones for which a combination of the above symmetries 
is conformal \cite{[symbh]}. Moreover, except for $V=0$, 
the interaction with conformal 
matter destroys the invariance under the other symmetries.

\section{Area-preserving couplings}

Conformal invariance provides   
little room to move in: it does 
not serve for generic 
potentials but only very particular ones. 
Therefore, to find solvable models for arbitrary potentials,  
we should go beyond conformal symmetry and consider interactions 
which are invariant under (some of) the generalized 
symmetries, $\delta_f$ and $\delta_2$,   
which have been described above. 

An important feature shared by $\delta_1$ and $\delta_2$ 
(but none of the other symmetries) 
is that they are area-preserving 
transformations -- that is, $g^{\mu\nu}\delta_{1,2}\>g_{\mu\nu}=0$. 
Therefore, if $S_M$ is invariant under area preserving 
transformations, the whole action $S_{GDG}$ will be 
invariant under $\delta_1$ and $\delta_2$ . 

Invariance under area-preserving transformations (APT)
requires that the traceless part of the energy-momentum tensor 
vanishes. Thus    

\be T_{\mu\nu}= \frac12 g_{\mu\nu}{T^\alpha}_\alpha\equiv 
\frac12 g_{\mu\nu}T\label{stT}\ee 
Hence, when the coupling is to area-preserving matter, 
the equations of motions (\ref{eqofmot}) take the form: 

\bea 
R+ V'(\phi)-T_\phi&=&0\nonumber\\
g_{\mu\nu}\square\phi -\frac12g_{\mu\nu}V(\phi)
- \nabla_\mu\nabla_\nu\phi-\frac12g_{\mu\nu}T&=&0\label{eqofmot2}\\
\left({E-L}\right)_A&=&0\nonumber 
\eea 

We can consider ${1\over2\pi} \int d^2x\sqrt{-{g}}
{V} ({\phi})$ to be part of the gravity-matter interaction term 
$S_M$. Therefore, without loss of generality and after 
a bit of algebra, we can write the equations of motion as follows 

\bea 
R&=&T_\phi\nonumber\\
\square\phi&=&T\label{eqofmotsimple}\\
\nabla_\mu\nabla_\nu \phi 
-\frac12g_{\mu\nu}\square\phi&=&0\nonumber\\
\left({E-L}\right)_A&=&0\nonumber
\eea 
The last-but-one equation implies that the vector  

\be k^\mu=\frac{\epsilon^{\mu\nu}}{\sqrt{-g}}\nabla_\nu\phi\ee 
satisfies the Killing equation $\nabla_{(\mu}k_{\nu)}=0$   
on shell. 

Now, invariance under diffeomorphisms of $S_M$ implies that 
on solutions of the equations of motion for the matter 
fields we have 

\be 0= T_\phi\nabla_\mu\phi + \nabla^\nu T_{\mu\nu}= 
T_\phi\nabla_\mu\phi + \nabla_\mu T\label{invdiff}\ee 
Hence, we also have  

\be k^\mu \nabla_\mu T= 0 \label{kT}\ee 
Therefore, on solutions of the equations of motion
for the matter fields, $T$ can be written as a function of the 
dilaton field:  

\be T= T(\phi) \ee 
Moreover, Eq. (\ref{invdiff}) implies that 
$T_\phi\equiv T_\phi(\phi)$ and 

\be T_\phi(\phi)=-T^\prime(\phi)\ee 

Therefore, we have shown that, with respect to the dilaton-gravity  
sector, the interaction with area-preserving matter 
simply amounts to a modification of 
the dilaton potential $V(\phi)\rightarrow \widetilde{V}(\phi)=
V(\phi)-T$. 

\subsection{Particular cases}

For a theory to be invariant under area preserving 
symmetries, it is a necessary and sufficient 
condition that its action depends on $g_{\mu\nu}$ exclusively 
through the measure $\sqrt{-g}$ \cite{[Zhao]}. 

Invariance under all area-preserving transformations  
imposes a severe restriction on the construction of 
interaction terms. However, in two dimensions, there is  
a large class of interactions,  notably Yang-Mills 
and generalized gauge theories,  
which fulfils this requirement \cite{[Zhao]}. 
In fact, in two dimensions any interaction 
term where the metric only raises antisymmetrised indices  
is invariant under area-preserving transformations. 
For instance, for a Yang-Mills theory in two dimensions, 
we have $F_{\mu\nu}=\epsilon_{\mu\nu}\widetilde{F}$. 
Therefore, the action has the form:

\be S_{Y-M}= \int \d^2 x \sqrt{-g}\hbox{Tr}F^{\mu\nu}F_{\mu\nu}
=-2\int \d^2 x \frac{\hbox{Tr}\widetilde{F}^2}{\sqrt{-g}}\ee
However, a minimal coupling of gauge 
fields to matter fields in an invariant way, 
is not possible.  

First of all, let us consider an interaction term which 
is invariant under area-preserving transformations 
of the metric and which does not depend on the dilaton 
field $\phi$. Then, we have 

\be \nabla_\nu T =0\>\Longrightarrow T=\hbox{constant}
\label{Tconstant}\ee  

In this case, therefore, in as far as the gravity sector is 
concerned, the interaction with matter  simply amounts  
to a constant shift of the potential 

\be V\longrightarrow \widetilde{V}=V - Q \label{shiftV1}\ee 
where $Q={T^\alpha}_\alpha=$ constant. 

Consider now a coupling 

\be S_M = {1\over2\pi} \int d^2x\sqrt{-g} 
{{\cal L}}_M\ee  
with 

\be {\cal L}_M=W(\phi)(\sqrt{-g})^s \widetilde{L}\ee 
where $\widetilde{L}$ depends only on the matter fields. 
Most of the couplings with gauge fields which have been 
considered in the literature are of this form. For 
instance, in Ref. \cite{[abelian]} a coupling of this form 
with an abelian gauge field is considered and in  
Ref. \cite{[yangmills]} a similar coupling with a Yang-Mills 
field is studied. 

In this case, invariance under diffeomorphisms implies 

\be W(\phi)^{\frac{s}{1+s}}(\sqrt{-g})^s \widetilde{L}
 =Q=\hbox{constant}\ee
Therefore, the shift in the potential is non-constant 
now but given by 

\be V\>\longrightarrow \widetilde{V}=V 
-(1+s)Q W(\phi)^{\frac1{1+s}}\ee

As a matter of fact, as precise a  
result can also be obtained with any 
interaction term in which the measure $\sqrt{-g}$ and the 
dilaton field $\phi$ appear only through a product 
$U(\phi)\sqrt{-g}$. Namely,  let $S_M$  
be a gravity-matter interaction term such that the dependence of 
its scalar density $\sqrt{-g} {{\cal L}}_M$  with respect to 
the measure $\sqrt{-g}$ and the dilaton field $\phi$ is 
of the form 

\be \sqrt{-g} {{\cal L}}_M \equiv \left(\sqrt{-g} 
{{\cal L}}_M\right)\>\left(y,\>A\right)\ee 
with $y = \sqrt{-g}U(\phi)$. Then, on matter fields which obey 
the equations of motions, we have      

\be \frac{\delta  (\sqrt{-g}{{\cal L}}_M) }{\delta y}
=Q=\hbox{constant}\ee
Therefore, the dynamics of the dilaton-gravity sector is the same as 
that without matter fields but with a different potential, 
$\widetilde{V}$, with 

\be \widetilde{V}(\phi) = V(\phi)-QU(\phi)\ee 

\section{Coupling by means of an invariant metric}

Another way of producing symmetric matter-gravity 
interactions is 
by considering couplings which involve only an invariant 
metric $\bar{g}_{\mu\nu}$. That is, let $\bar{g}_{\mu\nu}$ be a metric which 
is invariant under a transformation $\delta$ which, in turn, is a symmetry 
of $S_V$. Then $\delta$ is also  a symmetry of the action  

\be S=S_V + \int d^2x \sqrt{-\bar{g}}{\cal L}(\bar{g}_{\mu\nu},
f^A)\ee 
However, to demand strict invariance of the metric is in fact   
too restrictive a requirement. 
A more relaxed but sufficient condition is to require 
$\sqrt{-\bar{g}}{\cal L}(\bar{g},f^A)$ to be invariant. 
Consider, for instance a minimal coupling 

\be \sqrt{-\bar{g}}{\cal L} = \sqrt{-\bar{g}}\bar{g}^{\mu\nu}
\bar\nabla_\mu f\bar\nabla_\nu f\ee 
where $f$ is a scalar field. 
This interaction term is invariant under $\delta$ if  
$\bar{g}_{\mu\nu}$ is conformally invariant  

\be \delta \bar{g}_{\mu\nu} = K\bar{g}_{\mu\nu}\ee 
with arbitrary scalar quantity $K$. 

We shall restrict ourselves to metrics of the form 

\be \bar{g}_{\mu\nu}= A g_{\mu\nu} + B\nabla_\mu\phi\nabla_\nu\phi\ee 
with $A= A((\nabla\phi)^2,\phi)$ and $B=B((\nabla\phi)^2,\phi)$. 
We have 

\bea \det\g &=& A^2\det g(1+\frac{B}A(\nabla\phi)^2)\nonumber\\
\g^{\mu\nu} &=& \frac1A\left(g^{\mu\nu} -
\frac{\nabla^\mu\phi\nabla^\nu\phi}{\frac{A}B 
+(\nabla\phi)^2}\right)\\
\sqrt{-g}\g^{\mu\nu}&=&\sqrt{-g}\left(1+
\frac{B}{A}(\nabla\phi)^2\right)^\frac12\left(g^{\mu\nu} -
\frac{\nabla^\mu\phi\nabla^\nu\phi}{\frac{A}B +(\nabla\phi)^2}\right)
\nonumber\eea 
Due to the following transformation properties 

\bea \delta_f E = \frac\epsilon2f(E),&& \qquad \delta_f \phi =0\\
\delta_2 E =0,&&\qquad \delta_2\phi=\epsilon\eea
it is best to consider  
$A=A(E,\phi)$ and $B=B(E,\phi)$.

Invariance under $\delta_f$ and $\delta_2$ requires, respectively, 

\bea
\frac12f(E)\frac{\delta A}{\delta E} 
+\frac{f(E)A}{2E +J} + Af'(E)&=&K_f A\nonumber\\
\frac12f(E)\frac{\delta B}{\delta E} -2\frac{f(E)A}{(2E+J)^2}
- \frac{f'(E)A}{2E+J}&=&K_f B\nonumber\eea

\bea
\frac{\delta A}{\delta \phi}  
+\frac{AV}{2E+J}&=&K_2 A\nonumber\\
\frac{\delta B}{\delta\phi} -2\frac{AV}{(2E+J)^2}
&=&K_2 B\nonumber.
\eea

With these premises, it is not difficult to show  
that the most general conformally invariant or 
strictly invariant metrics $\g_{\mu\nu}$ (and invariant 
minimal couplings $\sqrt{-\g}\g^{\mu\nu}$) 
are of the following form ($A_f$, $A_2$, $A$, $\lambda_f$ and $\lambda_2$ 
are functions of their arguments and $\lambda$ is a constant):

\noindent -- Conformally invariant under $\delta_f$:  

\be \g_{\mu\nu}=A_f(E,\phi)\left(g_{\mu\nu} + 
\frac{\lambda_f^2(\phi) (\nabla\phi)^4f^2(E)-1}{\npdos} 
\nabla_\mu\phi\nabla_\nu\phi\right)\ee 

\noindent -- Strictly invariant 
under $\delta_f$:

\be \g_{\mu\nu}=\frac{A_f(\phi)}{\npdos f^2(E)}\left(g_{\mu\nu} + 
 \frac{\lambda^2_f(\phi)(\nabla\phi)^4 f^2(E)-1}{\npdos} 
\nabla_\mu\phi\nabla_\nu\phi\right)\ee 

\noindent -- Conformally 
invariant under $\delta_2$:

\be \g_{\mu\nu}=A_2(E,\phi)\left(g_{\mu\nu} + 
\frac{\lambda^2_2(E) (\nabla\phi)^4-1}{\npdos} 
\nabla_\mu\phi\nabla_\nu\phi\right)\ee 

\noindent -- Strictly 
invariant under $\delta_2$:

\be \g_{\mu\nu}=\frac{A_2(E)}\npdos\left(g_{\mu\nu} + 
\frac{\lambda^2_2(E) (\nabla\phi)^4-1}{\npdos} 
\nabla_\mu\phi\nabla_\nu\phi\right)\ee 

\noindent -- Conformally 
invariant under $\delta_f$ and $\delta_2$:

\be \g_{\mu\nu}=A(E,\phi)\left(g_{\mu\nu} + 
\frac{\lambda^2(\nabla\phi)^4f^2(E)-1}{\npdos} 
\nabla_\mu\phi\nabla_\nu\phi\right)\ee 

\noindent -- Strictly 
invariant under $\delta_f$ and $\delta_2$:

\be \g_{\mu\nu}=\frac1{\npdos f^2(E)}\left(g_{\mu\nu} + 
\frac{\lambda^2(\nabla\phi)^4f^2(E)-1}{\npdos} 
\nabla_\mu\phi\nabla_\nu\phi\right)\ee 

\noindent -- The most general minimal couplings 
which are invariant under $\delta_f$ are of the form 

\be \sqrt{-\g}\g^{\mu\nu} =\sqrt{-g}\lambda_f(\phi)\npdos f(E)
\left(g^{\mu\nu} - \frac{\lambda_f^2(\phi)(\nabla\phi)^4f^2(E) -1}
{\lambda_f^2(\phi)(\nabla\phi)^6f^2(E)}\nabla^\mu\phi\nabla^\nu\phi\right)
\ee

\noindent -- The most general minimal couplings 
which are invariant 
under $\delta_2$ are of the form 

\be \sqrt{-\g}\g^{\mu\nu} =\sqrt{-g}\lambda_2(E)\npdos 
\left(g^{\mu\nu} - \frac{\lambda_2^2(E)(\nabla\phi)^4 -1}
{\lambda_2^2(E)(\nabla\phi)^6f^2(E)}
\nabla^\mu\phi\nabla^\nu\phi\right)
\ee

\noindent -- The most general minimal couplings which are invariant 
under $\delta_f$ and $\delta_2$ are of the form 

\be \sqrt{-\g}\g^{\mu\nu} =\sqrt{-g}\npdos f(E) 
\left(g^{\mu\nu} - \frac{\lambda^2(\nabla\phi)^4 -1}
{\lambda^2(\nabla\phi)^6f^2(E)}
\nabla^\mu\phi\nabla^\nu\phi\right)
\ee

Therefore, metrics and conformal couplings 
exist which are invariant under both $\delta_f$ and $\delta_2$, 
for any function $f=f(E)$. 
However, it also follows that 
no metric or conformal coupling can be found which is 
invariant under $\delta_f$ and $\delta_g$  unless $f$ and $g$  
are  proportional to one another. 

The metrics which are conformally invariant under 
$\delta_1$ and $\delta_2$ turn out to be of the form 

\be \bar{g}_{\mu\nu}= A(E,\phi)\left(g_{\mu\nu} + 
\frac{\lambda^2(\nabla\phi)^4-1}{\npdos}
\nabla_\mu\phi\nabla_\nu\phi\right)\ee 
Strict invariance under $\delta_1$ 
and $\delta_2$ requires $A\propto\frac1\npdos$. 

Moreover, the only minimal couplings to a scalar 
field which are invariant under $\delta_1$ and $\delta_2$ 
are proportional to   

\be \sqrt{-\bar{g}}(\bar{\nabla} f)^2=
\sqrt{-g}(\nabla\phi)^2\left(g^{\mu\nu} - 
\frac{\lambda^2(\nabla\phi)^4-
1}{\lambda^2(\nabla\phi)^6}\nabla^\mu\phi\nabla^\nu\phi\right)
\nabla_\mu f\nabla_\nu f\label{confcoupling}\ee 
Thus,  save for a constant parameter,  
there is only one minimal coupling which is invariant under 
$\delta_1$ and $\delta_2$. 

\section{Discussion}

Consider again the CGHS and the exponential models coupled to 
conformal matter. Both these models are invariant   
under a conformal transformation but not the same 
transformation. The CGHS model coupled to conformal matter 
is invariant under $\delta_2 -4\lambda^2\delta_1$, whereas 
the exponential model is invariant with respect to  
$\delta_2 +2\beta \delta_E$. Moreover, the coupling to matter does not 
preserve any of the other symmetries that have been discussed in 
the present paper. 
As has been shown in Ref. \cite{[Kazama]}, invariance 
of Sigma models, as the CGHS or 
exponential models minimally coupled to matter,  
implies that these models contain a free field as well 
as a field which obeys a Liouville equation. However, 
as the Liouville equation does not appear  
as the conservation equation of a Noether current, this  
equation is not directly related to an invariance 
of the theory. Therefore, as there is 
not enough symmetry, no quantum solvability should be expected.  
In fact, the quantum nature of Liouville theory 
has remained elusive (see, for instance, Ref. \cite{[Kazama2]}) 
and much the same can be said about the quantum nature of the 
CGHS model (see, for instance, Ref. \cite{[Jackiw2]}). 

Unlike the CGHS and the exponential models, which are 
invariant with respect to one symmetry only (in this respect, 
the model with $V=0$ is somewhat special as it is  
invariant under two symmetries),    
we have constructed ``minimal'' couplings which 
are invariant under two symmetries. Therefore we expect 
that this additional invariance of the models  
will imply quantum as well as classical solvability.  
The analysis in Ref. \cite{[Kazama]} does not apply to our 
models and a different analysis should be made. 
A detailed discussion of this question 
deserves a separate study. 

Finally, we would like to mention that,  
for more than two decades now, it has been known that 
Einstein's gravity, when restricted to metrics 
with two {\it commuting} Killing vector fields, acquires a large number of 
nonabelian symmetries, the so-called Geroch 
group \cite{[Geroch]}. It is clear that our results 
may have some relationship 
with, or be a generalization of, the Geroch's group. 
We hope to stablish that relationship and communicate it   
in a future publication. 

\section*{Acknowledgements}
The author is grateful to J. Cruz, J. Navarro-Salas and 
C.F. Talavera for helpful discussions, and to 
G. Mena-Marug\'an for calling the author's attention to the Geroch group. 
He acknowledges the Spanish MEC, CSIC and IMAFF (Madrid)
for a research contract.

\end{document}